\documentstyle[11pt,epsfig]{article}

\newlength{\dinwidth}
\newlength{\dinmargin}
\newcommand{\PLB}[3]{Phys.~Lett.\ {\bf B#1} ({#3}) {#2}}
\newcommand{\NPB}[3]{Nucl.~Phys.\ {\bf B#1} ({#3}) {#2}}
\newcommand{\PRD}[3]{Phys.~Rev.\ {\bf D#1} ({#3}) {#2}}

\newcommand{\ZPC}[3]{Z.~Phys.\ {\bf  C#1} ({#3}) {#2}}

\newcommand{\e}{\mathrm{e}}

\newcommand{\alps}{\alpha_s}

\newcommand{\deta}{\Delta \eta}

\def\lapproxeq{\lower .7ex\hbox{$\;\stackrel{\textstyle<}{\sim}\;$}}
\def\gapproxeq{\lower .7ex\hbox{$\;\stackrel{\textstyle>}{\sim}\;$}}

\begin{document}

\sloppy

\begin{titlepage}

  \begin{flushright}
    CERN--TH/99--393\\
    MC-TH-99/17 \\
    December 1999
  \end{flushright}

\begin{center}
    
\vskip 10mm {\Large\bf\boldmath Diffractive production of high $p_t$ photons 
at a future linear collider} \vskip 15mm

    {\large Norman Evanson}\\
    $^*$Dept.~of Physics and Astronomy, University of Manchester\\
    Manchester M13 9PL, England\\
    \vskip 10mm

    {\large Jeff Forshaw}\footnote{On leave of absence from $^*$}\\ 
    Theory Division, CERN \\
    1211 Geneva 23, Switzerland\\
    \vskip 10mm

\end{center}

\vspace*{0.4cm}
\begin{abstract}
We examine the prospects for studying the diffractive production of high
$p_t$ photons in the process $\gamma \gamma \to \gamma X$ at a future 
linear collider operating in both $ee$ and $\gamma \gamma$ modes. The
high luminosity associated with a linear collider make it the ideal place
to measure this process.
\end{abstract}

\end{titlepage}

\section{Introduction}
In the pursuit to understand diffraction in strong interactions it is
sensible to focus on those rapidity gap processes that we are in principle
able to compute reliably, i.e. using QCD perturbation theory. Of all such
processes, diffractive high $p_t$ photon production stands out as the one
most accessible to perturbation theory. The process can already be studied
at HERA via $\gamma p \to \gamma X$ where the final state photon is well
separated in rapidity from the debris of the proton, $X$ 
\cite{Ivanov1,Ivanov2,Evanson,Cox}. Note that it is not necessary to measure
the system $X$ and that this enhances the reach in rapidity significantly.
We shall show that the equivalent process can be studied at proposed 
future high energy $e^+ e^-$ and $\gamma \gamma$ 
colliders with much higher event rates. Note that the process 
$\gamma \gamma \to \gamma \gamma$ is in principle also possible. However, 
the rates for this process are probably too low even for a future linear 
collider and we do not consider it further. We also remark that the current 
LEP collider is not able to measure either of these processes due to 
insufficient luminosity. For a much more detailed account of most of the 
results presented here, we refer to \cite{thesis}.

Theoretical interest in the process $\gamma \gamma \to \gamma X$ dates back 
to the work of \cite{Ginzburg} where calculations were performed in fixed 
order perturbation theory and to lowest order in $\alpha_s$. Recent work has 
extended this calculation to sum all leading logarithms in energy, for real 
incoming photons \cite{Ivanov2} and for real and virtual incoming photons 
\cite{Evanson,thesis}. The cross-section for relevant hard subprocess,
$\gamma q \to \gamma q$, can be written
\begin{equation}
\frac{d \sigma_{\gamma q}}{dp_t^2} 
\approx \frac{1}{16 \pi \hat{s}^2} \left| A_{++} \right|^2
\end{equation}
and we have ignored a small contribution that flips the helicity of the
incoming photon. The photon-quark CM energy is given by $\hat{s}$. To leading
logarithmic accuracy \cite{Ivanov2,Evanson}
\begin{equation}
A_{++} = i \alpha \alpha_s^2 \sum_q e_q^2 \frac{\pi}{6} \frac{\hat{s}}{p_t^2}
\int_{-\infty}^{\infty} \frac{d\nu}{1+\nu^2} \frac{\nu^2}{(\nu^2+1/4)^2}
\frac{{\rm tanh} \pi \nu}{\pi \nu} F(\nu) \ \e^{z_0 \chi(\nu)}
\end{equation}
where
\begin{equation}
z_0 \equiv \frac{3 \alpha_s}{\pi} \log \frac{\hat{s}}{p_t^2},
\end{equation}
$\chi(\nu) = 2(\Psi(1) - {\rm Re}\Psi(1/2+i \nu))$ is the BFKL eigenfunction
\cite{BFKL}, $F(\nu) = 2(11+12 \nu^2)$ for on-shell photons and there is a 
sum over the quark charges squared, $e_q^2$. The full photon-photon 
cross-section is obtained after multiplying by the photon parton density 
functions: 
\begin{equation}
\frac{d \sigma_{\gamma \gamma}}{dx dp_t^2} = 
\left[ \frac{81}{16} g(x,\mu) + \Sigma(x,\mu) 
\right] \frac{d \sigma_{\gamma q}}{dp_t^2}
\end{equation}
and we take the factorization scale $\mu = p_t$. Throughout we take
$\alps = 0.2$, as indicated by HERA and Tevatron data \cite{alps}.

The larger $z_0$, the larger the rapidity gap between the outgoing photon and
the system $X$. Since $$ z_0 \approx \frac{3 \alps}{\pi} \deta. $$
For $z_0 \gapproxeq 0.5$ one begins to access the most 
interesting Regge region \cite{Ryskin}. 

\section{\boldmath Results: $e^+ e^-$ mode}

Since it is not possible to measure this process at LEP, we turn our 
attention immediately 
to a future linear collider operating in $e^+ e^-$ mode and at 
$\surd s = 500$ GeV and $\surd s = 1$ TeV. We impose three different cuts on 
the minimum angle ($\theta_{\gamma}$) of the emitted photon and make a cut 
to ensure that its energy is above 5 GeV. Our results are insensitive 
to this photon energy cut since we also impose a cut on the subprocess 
centre-of-mass energy: $\hat{s} > 10^3$ GeV$^2$. This cut is not easy to 
implement experimentally, but is essentially related to the size of the 
final state rapidity gap. Making this cut ensures that $z_0$ is large, i.e. 
the photon and the system $X$ are well separated.
We also make an anti-tag cut on the outoing electron and positron, 
i.e. we insist that they be emitted at angles less than 100 mrad. When
quoting event rates, we assume a total integrated luminosity of 50~fb$^{-1}$. 
In all cases, we show the cross-section integrated over the $p_t$ 
of the emitted photon subject to the constraint 1~GeV~$<~p_t~<~10$~GeV and
over the invariant mass of the system $X$ subject to $M_X < 10$ GeV.
The photon flux of \cite{Frixione} and the photon
parton density functions of \cite{GRS} were used. Our 
results are summarized in Table~\ref{tab:NLC}.

\begin{table}
\begin{center}
\begin{tabular}{|c|c|c|}\hline 
\multicolumn{3}{|c|}{$\int dt~{\cal L}_{ee} = 50$ fb$^{-1}$} \\ \hline
$\theta_{\gamma}$ mrad & $\sigma$ pb  & Events\\ \hline
\multicolumn{3}{|c|}{$\sqrt{s_{ee}}=500$ GeV} \\ \hline    
100    & 0.14   & 6.9k \\
200    & 0.027  & 1.4k \\
300    & 0.0066 & 0.3k \\ \hline
\multicolumn{3}{|c|}{$\sqrt{s_{ee}}=1$ TeV} \\ \hline
100    & 0.41   & 21k \\
200    & 0.088  & 4.4k \\
300    & 0.021  & 1.0k \\ \hline
\end{tabular}
\caption{The integrated cross-section and number of events for 
$ee\to ee\gamma X$ at a future linear collider.}
\label{tab:NLC} 
\end{center}
\end{table}

The range of $z_0$ values accessible to a future linear collider are
restricted by the requirement that the photon appear in the detector. 
However, one does gain by going to higher beam energies since the dissociated
system $X$ does not need to be seen. In this sense optimal configurations
involve a relatively soft photon colliding with a hard photon. For the 
scenarios presented in the table, we find $0.4 < z_0 < 1.8$ which is well into
the region of interest. Due to the softness of the bremsstrahlung photon
spectrum one gains appreciably in rate by progressing to higher beam
energies. Note the very strong dependence upon the minimum angle of the
detected photon.  

\section{\boldmath Results: $\gamma \gamma$ mode}

Next we turn to the photon collider. For the flux of photons we use 
typical parameters \cite{Sitges}: $x=4.8$, $P_c = -1$ (i.e. negative 
helicity laser photons) and $2 \lambda = 1$ (i.e. 100\% longitudinally 
polarized electron and positron beams). The choice $2 \lambda P_c = -1$ 
leads to a hard photon spectrum:
\begin{equation}
\int_{0.65}^{z_{max}} dz \frac{d {\cal L}_{\gamma \gamma}}{dz}
= 30 \%
\end{equation}
where $z_{max} = x/(x+1)$, $z = W_{\gamma \gamma}/\surd s$ ($\sqrt{s}/2$ is 
the electron beam energy) and the integral down to $z=0$ is defined to give 
unity. We use the photon flux of \cite{Telnov} and normalize it to obtain the
$e^+ e^-$ cross-section using the ``rule of thumb''
${\cal L}_{\gamma \gamma}(z>0.65) \approx 0.15~{\cal L}_{ee}$ \cite{Sitges}.
In particular the $e^+ e^-$ rate is determined using
\begin{equation}
\sigma_{ee} = \frac{0.15}{0.30} 
\int_0^{z_{max}^2} d\tau \int_{\tau/z_{max}}^{z_{max}}
dy \ f(y,\tau) \ \sigma_{\gamma \gamma}
\end{equation}
where
\begin{equation}
\frac{d {\cal L}_{\gamma \gamma}}{d \tau} = \int_{\tau/z_{max}}^{z_{max}}
dy \ f(y,\tau)
\end{equation}
and $\tau = z^2$.  We are able to use a single photon luminosity function
since the $\gamma \gamma$ cross-section is independent of the photon
helicity (for transverse photons).

Table \ref{tab:CC} shows our results for the photon collider, assuming
the same cuts as in the previous section.

\begin{table}
\begin{center}
\begin{tabular}{|c|c|c|}\hline 
\multicolumn{3}{|c|}{$\int dt~{\cal L}_{ee} = 50$ fb$^{-1}$} \\ \hline
$\theta_{\gamma}$ mrad & $\sigma$ pb  & Events\\ \hline
\multicolumn{3}{|c|}{$\sqrt{s_{ee}}=500$ GeV} \\ \hline    
100    & 1.32   & 66k \\
200    & 0.20   & 10k \\
300    & 0.045  & 2.3k \\ \hline
\multicolumn{3}{|c|}{$\sqrt{s_{ee}}=1$ TeV} \\ \hline
100    & 1.40   & 70k \\
200    & 0.22   & 11k \\
300    & 0.048  & 2.4k \\ \hline
\end{tabular}
\caption{The integrated cross-section and number of events for 
$ee\to ee\gamma X$ at a future photon collider.}
\label{tab:CC} 
\end{center}
\end{table}

The $z_0$ range for the photon collider is similar to that for the $e^+ e^-$
collider, i.e. for the events in Table \ref{tab:CC} $0.4 < z_0 < 1.8$.
For the photon collider, the photon spectrum is not soft. This means that
ultimately the rate depletes as the beam energy increases. This can
be seen in the table where the 500 GeV and 1 TeV rates are similar. Going to
higher beam energies leads to a slow reduction in rate (however the typical
$z_0$ range does still move to higher values). For example, at a 5 TeV collider
the cross-section is 1.4 pb and the $z_0$ range is $0.5 \sim 2$ 
($\theta_{\gamma} = 0.1$). The drop in rate as one increases the detected
photon angle is even more dramatic than in the $e^+ e^-$ mode and is a
consequence of the harder photon spectrum. 

One can imagine further improving the capabilities of the photon collider by 
arranging to use one soft photon spectrum and one harder spectrum. The
harder photon dissociates into a very forward system whilst the softer 
photon is easier to scatter into the detector. One way to do this would be
to operate the photon collider with $2 P_c \lambda = +1$ to produce the 
softer spectrum and $2 P_c \lambda = -1$ for the harder spectrum. Even 
better would be to operate in the $e \gamma$ mode where a soft bremsstrahlung 
photon is made to collide with a hard Compton photon.  

\section{Summary}

A linear $e^+ e^-$ collider operating at 500 GeV and beyond is the ideal 
place to study the diffractive production of high $p_t$ photons via 
$\gamma \gamma \to \gamma X$. It will have the capacity to provide
very important information in abundance on the short distance domain of 
large rapidity gap physics. Operating in the photon collider mode offers the 
opportunity to further enhance the rate. In all cases, the rate grows 
rapidly with decreasing angle of the detected photon.

This short note has focussed on the production of high $p_t$ photons. Also
of tremendous interest is the production of high $p_t$ vector particles
in general, e.g. $\rho, \omega, \phi$ and $J/\Psi$. These processes would
also occur in abundance at a future linear collider.

\section*{Acknowledgements}
Thanks to Albert de Roeck and Stefan Soldner-Rembold for helpful discussion
and to Georgi Jikia for providing the code to compute the photon luminosity for
the photon collider.

\end{document}